\documentclass[10pt,sigconf,letterpaper,nonacm]{acmart}
\usepackage[T1]{fontenc}
\usepackage{graphicx}
\usepackage{xcolor}
\usepackage{blindtext}

\newcommand{\paul}[1]{\ignorespaces}
\newcommand{\fb}[1]{\ignorespaces}
\newcommand{\nick}[1]{\ignorespaces}

\newcounter{packednmbr}
\newenvironment{packedenumerate}{\begin{list}{\thepackednmbr.}{\usecounter{packednmbr}\setlength{\itemsep}{0.5pt}\addtolength{\labelwidth}{-4pt}\setlength{\leftmargin}{2ex}\setlength{\listparindent}{\parindent}\setlength{\parsep}{1pt}\setlength{\topsep}{0pt}}}{\end{list}}
\newenvironment{packeditemize}{\begin{list}{$\bullet$}{\setlength{\itemsep}{0.5pt}\addtolength{\labelwidth}{-4pt}\setlength{\leftmargin}{2ex}\setlength{\listparindent}{\parindent}\setlength{\parsep}{1pt}\setlength{\topsep}{2pt}}}{\end{list}}

\newcommand{\tightcaption}[1]{\vspace{-0.4cm}\caption{{\normalfont{\textbf{{#1}}}}}\vspace{-0.3cm}}

\newcommand{\eg}{{\it e.g.,}\xspace}
\newcommand{\ie}{{\it i.e.,}\xspace}

\newcommand{\mypara}[1]{\vspace{0.05cm}\noindent{\bf {#1}:}~}

\definecolor{backcolour}{rgb}{0.96,0.96,0.96}
\definecolor{codegray}{rgb}{0.5,0.5,0.5}
\definecolor{deepblue}{rgb}{0,0,0.6}
\definecolor{deepred}{rgb}{0.6,0,0}
\definecolor{deepgreen}{rgb}{0,0.5,0}

\usepackage{listings}
\usepackage{xcolor}

\lstdefinestyle{mypython}{
    language=Python,
    basicstyle=\ttfamily\small,
    keywordstyle=\color{blue},
    stringstyle=\color{teal},
    commentstyle=\color{gray},
    breaklines=true,
    frame=single,
    showstringspaces=false
}

\usepackage{float}
\floatstyle{ruled}
\newfloat{listing}{htbp}{lop}
\floatname{listing}{Listing}

\lstdefinestyle{mybash}{
    language=Bash,                   
    basicstyle=\ttfamily\small,      
    keywordstyle=\color{blue},       
    stringstyle=\color{teal},        
    commentstyle=\color{gray},       
    breaklines=true,                 
    frame=single,                     
    showstringspaces=false,           
    morekeywords={sudo,ssh,scp,rm,mkdir,echo,kill} 
}
\begin{document}
\title{Characterizing the Impact of Active Queue Management on Speed Test Measurements}
%
%


\author{Siddhant Ray}
\affiliation{
  \institution{University of Chicago}
  \city{}
  \state{}
  \country{}
}

\author{Taveesh Sharma}
\affiliation{
  \institution{University of Chicago}
  \city{}
  \state{}
  \country{}
}

\author{Jonatas Marques}
\affiliation{
  \institution{University of Chicago}
  \city{}
  \state{}
  \country{}
}

\author{Paul Schmitt}
\affiliation{
  \institution{Cal Poly}
  \city{}
  \state{}
  \country{}
}

\author{Francesco Bronzino}
\affiliation{
  \institution{ENS Lyon}
  \city{}
  \country{}
}

\author{Nick Feamster}
\affiliation{
  \institution{University of Chicago}
  \city{}
  \state{}
  \country{}
}

\renewcommand{\shortauthors}{Ray et al.}

\begin{abstract}

Present day speed test tools measure peak throughput, but often fail to capture the user-perceived responsiveness of a network connection under load. 
Recently, platforms such as NDT, Ookla Speedtest and Cloudflare Speed Test have introduced metrics such as ``latency under load'' and ``real-time throughput variation'' to fill this gap. 
Yet, the sensitivity of these metrics to basic network configurations such as Active Queue Management (AQM) remains poorly understood. 
In this work, we conduct an empirical study of the impact of AQM on speed test measurements in a laboratory setting. 
Using controlled experiments, we compare the distribution of throughput and latency under different load measurements across different AQM schemes, 
including CoDel, FQ-CoDel and Stochastic Fair Queuing (SFQ).
Compared to a standard drop-tail baseline, we find that measurements have high variance across AQM schemes and load conditions.
These results highlight the critical role of AQM in shaping how emerging latency metrics should be interpreted, and underscore the need for better calibration 
of speed test platforms before their results are used to guide policy or regulatory outcomes.


\end{abstract}

\maketitle              

\section{Introduction}
\label{sec:intro}

Internet performance has historically been summarized using a single number: ``speed'' \cite{bauer2010understanding,midoglu2018monroenettestconfigurabletooldissecting,feamster2019internetspeedmeasurementcurrent}. 
Despite their widespread utility, traditional measurements that focus on peak bandwidth (\ie link capacity) fail to capture user-perceived Quality of Experience (QoE). For many interactive applications (e.g., remote collaboration, gaming, video conferencing), QoE is determined not by the maximum attainable rate but by the available bandwidth and latency under load, which conventional full-capacity speed tests do not accurately reflect.
However, the interpretation and use of these metrics has not been standardized. For instance, NDT \cite{ndt}, reports average throughput and some aggregated measurements of total bytes transferred during the test.
Ookla defines ``working latency'' as the increase in round-trip time (RTT) under load compared to the unloaded RTT, measured during a speed test \cite{CeroWRT_speedtests}. Apple uses a different metric, called round trips per minute (RPM) under load, which counts the number of round trips completed during a 
fixed time interval while the connection is saturated \cite{ietf-ippm-responsiveness-07}. Further, these tests have been known to discard outliers that often correspond to glitches that users 
typically notice during real-time applications such as video conferencing and streaming \cite{CeroWRT_speedtests}. As a result, users and regulators are left with incomplete pictures of what causes an Internet connection to be unresponsive, 
and how it can be mitigated. A central, unanswered question is how traditional metrics such as throughput and latency, and new metrics such as LUL, behave in the presence or absence of active queue management (AQM) algorithms such as FQ-CoDel, 
which were explicitly designed to maintain low latency under load \cite{hoilandbufferbloat}.

While AQM has been widely studied in the context of TCP performance, its impact on speed test measurements remains underexplored.
Given that AQM deployment is steadily increasing\footnote{Based on conversations with leading ISP vendors in the United States.}, understanding this relationship is crucial for both network operators and end-users to accurately assess and improve their internet experience.
In this paper, we investigate how the empirical distribution of modern speed test measurement results shifts when an AQM is deployed. Rather than reporting only typical throughput (e.g., mean) and latency values (e.g., $90^{th}$ percentile, median), we analyze full distributions: the tails, the spikes, and metrics similar to ``glitches per minute'' \cite{CeroWRT_speedtests} that are most relevant to real-time applications. Our goal is to empirically characterize the difference between unmanaged queues and AQM-enabled network paths, and to highlight how this difference is (or is not) reflected in widely deployed measurement platforms. By doing so, we aim to inform both test designers and network operators of the gaps between the status quo of Internet measurement and the actual experience of end-users.

In summary, our contributions are as follows:

\begin{packeditemize}
    \item We conduct a measurement study with popular speed test tools of (NDT)~\cite{ndt} and iperf3~\cite{iperf3}across varying AQM algorithms and network conditions 
    to analyze impact of such policies and link load on speed test measurements.
    \item We show that traditional speed test metrics such as averaged throughput and latency can mask significant variations in user-perceived speed and responsiveness and believe 
    that our findings should help inform the design of future speed test tools should consider network conditions and AQM policies to report more interpretable metrics.
\end{packeditemize}
\section{Background}
\label{sec:background}

\subsection{Active Queue Management (AQM)}

AQM techniques have been an active area of research and deployment for the past few decades to reduce latency and bufferbloat~\cite{bufferbloat} and to ensure fairness and coexistence 
among TCP flows on the shared network links. Traditional AQM algorithms have been built to run in conjunction with TCP congestion control algorithms, which rely only on packet loss as a signal for congestion.

To handle the bursty nature of TCP, these AQM techniques are equipped with large data buffers to prevent excessive packet drops due to these bursts. 
However, bufferbloat arises when queue length grows unbounded, specifically if the buffers are increasingly large, the packets end up being queued in much deeper queues, leading to excessive queuing latency.
latency can often build up as a result of individual bufferbloats at multiple routers on the network path. 

AQM techniques such as Random Early Detection~\cite{red}, CoDel~\cite{codel} and FQ-CoDel~\cite{fqcodel} have been
designed to reduce latency and bufferbloat, by actively managing the queue lengths by dropping packets before the queue becomes full.
Other techniques such as Stochastic Fair Queuing (SFQ)~\cite{sfq} and Deficit Round Robin (DRR)~\cite{drr} aim to provide
fairness among competing flows by ensuring equal bandwidth allocation.
AQM techniques such as Proportional Integral controller Enhanced (PIE)~\cite{pie}, CAKE~\cite{cake} and Low Latency, Low Loss, Scalable Throughput (L4S)~\cite{l4s} 
have been proposed to provide low latency and high throughput for modern applications such as video streaming and online gaming.

These recent advances in AQM suggest that integrating design principles bufferbloat and fairness (e.g., CAKE and L4S), which can yield significant performance improvements over other traditional AQM techniques.
CAKE emphasizes deployability in heterogeneous consumer environments by incorporating bandwidth shaping, per-flow and per-host fairness, DiffServ awareness, and optional TCP ACK filtering to mitigate asymmetry and NAT-induced complexity. 
L4S introduces a congestion control framework based on Explicit Congestion Notification (ECN) signaling and shallow queuing, enabling sub-millisecond latency and high throughput when paired with responsive transport protocols. 

These AQM techniques are being widely adopted in various network devices~\cite{10.1007/978-3-031-85960-1_10, ietf-tsvwg-l4sops-08}, including home routers, enterprise networks, and data centers, to improve network performance and user experience. However,
measuring the impact of AQM techniques by speed test measurement tools remains underexplored due to the lack of standardized methodologies and metrics to evaluate their overall performance.

\subsection{Speed Test Measurement Tools}

A variety of speed test measurement tools are commonly employed to assess end-to-end network performance. 
The Measurement Lab's Network Diagnostic Tool (NDT)~\cite{ndt,ndt7spec} is a widely deployed open platform that enables broadband quality evaluation through standardized TCP-based tests. NDT establishes a controlled client--server connection to measure achievable throughput, round-trip latency, and packet loss, and in its most recent version (NDT7) uses WebSocket transport for compatibility with modern browsers. In addition to providing a user-facing interface, NDT publishes all test results as open data, thereby supporting reproducible, large-scale research on global internet performance.

Ookla's speed test~\cite{ookla,ookla_method} is a commercial measurement service that conducts real-time assessments of download and upload speeds, latency, and jitter. It typically operates by initiating multiple parallel TCP flows between the client and one of its geographically distributed test servers, thereby approximating the aggregate throughput available to a user under normal usage conditions. The service automatically selects the closest or least-loaded server to minimize measurement bias, and results are recorded through an interactive interface widely deployed across web and mobile platforms.

Apple's speedtest~\cite{apple_speedtest} application is designed for iOS devices with a lightweight interface to evaluate network performance directly from mobile hardware. The tool measures download/upload rates and latency using a set of Apple-operated servers integrated into the broader iOS ecosystem. Leveraging native APIs and mobile-specific optimizations, the application captures performance metrics reflecting real-world mobile usage scenarios.

Fast.com~\cite{fast}, developed by Netflix, is a web-based measurement service that focuses primarily on download speed evaluation, reflecting the bandwidth available for streaming media. The test runs automatically when a user visits the site, initiating multiple parallel HTTP-based flows to Netflix-operated content servers. The service also provides optional measurements of upload speed and latency under load, offering a lightweight, consumer-oriented approach to performance testing of speed test measurement.

These tools collectively provide standardized yet distinct methodological approaches for assessing network quality, supporting both individual diagnostics and broader internet performance research.
However, they share a common limitation in that they do not measure latency under sustained load conditions.
Latency under load is crucial for characterizing user-visible network performance as it reveals queueing delays and congestion behavior. 
Existing speed test tools primarily report peak throughput and baseline RTT, which underrepresents performance during sustained load. 
and provide an incomplete view of network quality and systematically misestimate real-world experience.

\subsection{Related Work}

speed test measurement research: Several studies have explored methodologies and challenges in measuring broadband performance using diverse tools and datasets \cite{ndt_mlab_pam2010,ndt_mlab_sigcomm2018,broadband_pam2019,bottleneck_imc2020,ookla_sigmetrics2016,fast_imc2022,crowd_pam2021}.
The authors behind the M-Lab platform have provided an open, large-scale testbed for broadband diagnostics, enabling researchers to analyze throughput, latency, and access network bottlenecks across user populations \cite{ndt_mlab_pam2010,ndt_mlab_sigcomm2018}.
Follow-up studies have leveraged M-Lab data to understand end-to-end performance and characterize latency dynamics in last-mile access networks \cite{broadband_pam2019,bottleneck_imc2020}.
Other commercial systems such as Ookla's speed test have been examined for large-scale performance estimation and for highlighting biases in crowdsourced measurement datasets \cite{ookla_sigmetrics2016}.
More recent work has analyzed web-based speed test platforms such as Netflix's Fast.com aimed at identifying how test design influences user-visible metrics \cite{fast_imc2022}.
Complementary analyses of measurement bias and representativeness in crowdsourced platforms emphasize the need for better calibration to interpret speed test-derived metrics \cite{crowd_pam2021}.

\section{Methodology}
\label{sec:methodology}

In order to study the impact of active queue management (AQM) on speed test results we set up an isolated testbed.
Our testbed is composed of two System76 Meerkat 6 mini computer hosts connected together with a Turris Omnia open-source router.
Both Meerkat mini computers are configured with an Intel Core i5-1135G7 (2.4GHz) processor, 16GB of DDR4 RAM (3200 MHz), a 2.5GbE-capable Intel Ethernet Controller I225-LM, and Ubuntu 20.04.5 LTS (GNU/Linux 5.18.10-76051810-generic x86\_64).
The Omnia router is configured with Marvell Armada 385 (1.6GHz), 2GB of DDR3 RAM, five 1GbE LAN ports, one 1GbE WAN port, and TurrisOS 7.2.3 (GNU/Linux 5.15.148).

With this 1GbE-capable testbed, we perform a series of speed tests to collect results in many specific scenarios varying a number of options and conditions.
Namely, we test different (a) speedtest tools, (b) AQM algorithms, (c) throughput limits, (d) latency values, (e) role of burst shaping for maximum link utilization, and (f) effect of competing traffic.

Regarding speed testing, we focus on two tools: Measurement Lab Network Diagnostic Tool (NDT) (with TCP BBR) and iperf3 based tests for TCP.
We have chosen these two tools due to their scale and popularity, in terms of number of vantage points, size of datasets, and their usage in research and policy communities \cite{cdcc2022illustrating,crpa2019broadband,clark2021measurement,clark2024measurement,hoiland-jorgensen2016measuring,jiang2023mobile,lee2023analyzing,macmillan2023comparative,nabi2024red,paul2022characterizing,paul2022importance,sanchez-arias2023understanding,saxon2022what,sharma2024beyond,sundaresan2017challenges}.
For both speed test tools, we designate one of our hosts as the server, installing the necessary server-side application, and run tests pointing to the server from the other host using command-line client applications.
We configure the speed test clients to output verbose results and save all data for post-hoc analysis. Finally as NDT does not report
instantaneous throughput measurements, we also collect pcaps to measure instantaneous throughput values.

We evaluate three individual AQM algorithms: CoDel (Controlled Delay), FQ-CoDel (Fair/Flow Queue CoDel), and SFQ (Stochastic Fair Queuing).
We also run tests with the default queuing discipline, i.e., drop-tail, which we refer to as No AQM.
These algorithms were selected because they are directly available in standard Linux distributions, and were configured on the TurrisOS Linux testbed router using \texttt{tc qdisc} with default Linux parameters: CoDel with \texttt{target = 5 ms}, \texttt{interval = 100 ms}, and \texttt{limit = 1000} packets; FQ-CoDel with \texttt{target = 5 ms}, \texttt{interval = 100 ms}, \texttt{limit = 10240} packets, \texttt{flows = 1024}, and \texttt{quantum = 1514} bytes; SFQ with \texttt{divisor = 1024}, \texttt{limit = 127}, \texttt{depth = 127}, and \texttt{quantum = MTU}; and \texttt{pfifo} with a tail-drop queue whose limit equals the interface \texttt{txqueuelen}.

For throughput limits, we consider the values between 100 Mbps and 1Gbps in steps of 100 Mbps, representing popular modern residential speeds.
Similar to AQM algorithm, we configure rate limiting using \texttt{tc} and the Hierarchy Token Bucket algorithm in the router.
We set the rate parameter to the respective throughput limit and we evaluate the setups with and without burst shaping enabled.
When burst shaping is enabled, we set the \texttt{ceil} parameter to 1.6$\times$ the baserate, the \texttt{burst} parameter to 15KB, and the \texttt{cburst} parameter to 30KB.
The respective values for \texttt{ceil} were chosen based on the common practice of setting ceil = 1.2$\times$ to 2$\times$ rate,
making 1.6$\times$ a balanced heuristic between fairness and flexibility \cite{floyd2001tcp}. For the \texttt{burst} and \texttt{cburst} parameters, we follow the practices suggested in the \texttt{tc} manual \cite{tc_manpage},
which recommend setting these values to accommodate for few hundred milliseconds of bursty traffic. This allows sustaining TCP data transmission and avoiding jitter without getting excessively aggressive 
at the given shaping rate.

We also experiment with introducing artificial delay to the interface connecting hosts in order to emulate the typical communication latency on the Internet.
Initially, we configure the router interface to introduce such delay using the \texttt{tc netem} command.
However, in our testing we detect that the CPU-intensive nature of this algorithm saturates the router and makes performance highly unstable.
To address this, we allow for optionally adding the delay in the server with a \texttt{netem} discipline acting over the outgoing traffic using Linux \texttt{tc} and \texttt{IFB}
(Intermediate Functional Block) device. Since Linux \texttt{tc} \texttt{netem} can only shape outbound traffic, an IFB device is used to redirect incoming packets so they can be shaped as if they were outgoing.
However, we observed that adding such latency merely offsets the latency measurements by a constant factor, without affecting throughput or relative latency under load.
Therefore, in this paper we only present results without added latency.

Finally, we evaluate the effect of competing traffic in the path between client and server hosts.
We introduce competing traffic by running the \texttt{iperf3} tool at the same time as the speed tests.
We use TCP Cubic to add cross traffic as it is the default congestion control algorithm in Linux.
Our experimental setup consists of starting an (indefinitely-running) \texttt{iperf3} test 5 seconds before launching the speed test tool and 
once the speed test ends, we also terminate the \texttt{iperf3} test.
We allow the cross traffic to be capacity-seeking (we do not set a target rate) to use all available bandwidth.
\section{Speed Test Measurement Study Results}
\label{sec:results}

\begin{figure}[t]
    \centering
    \includegraphics[width=0.9\linewidth]{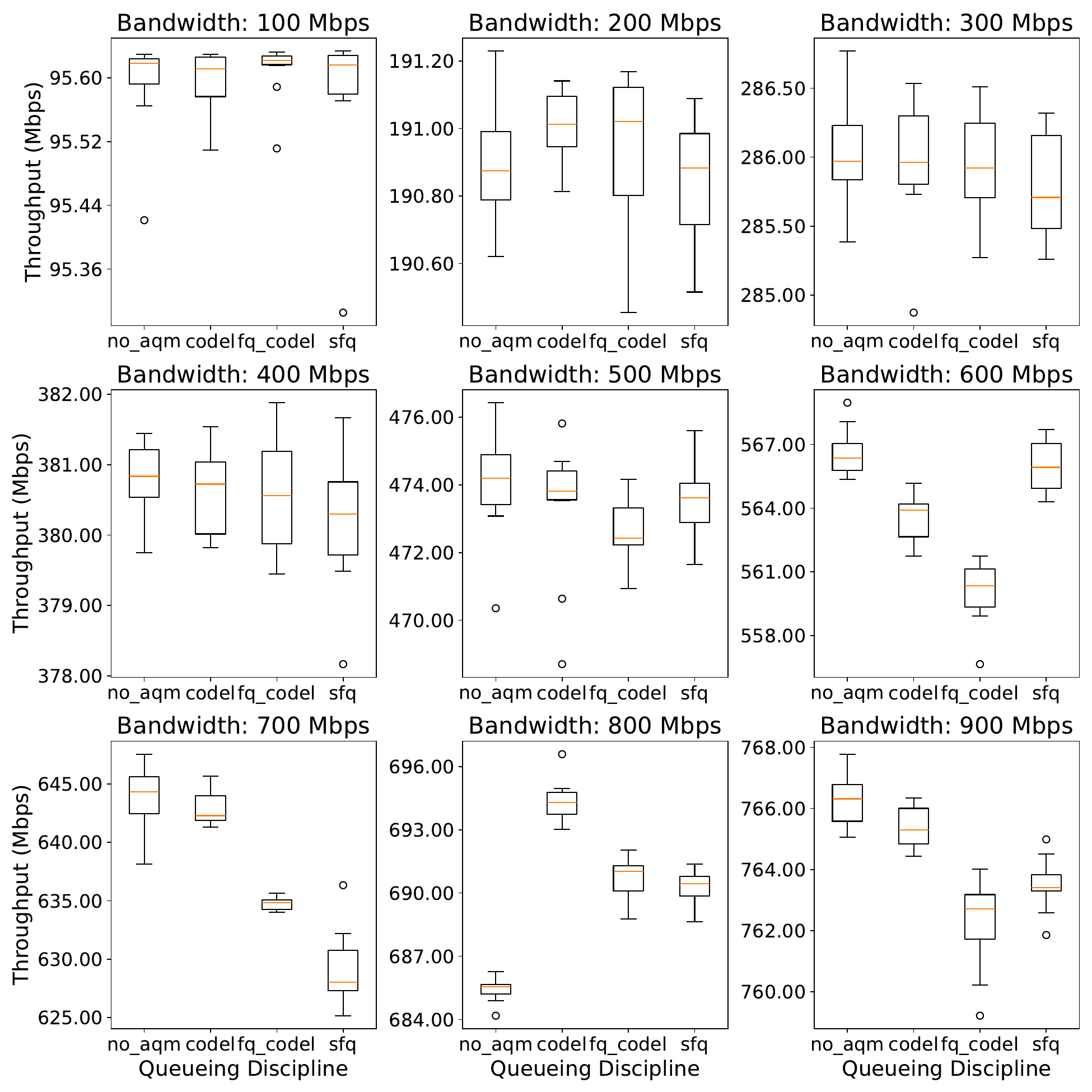}
    \tightcaption{Throughput measurements without burst shaping and without cross traffic has lower variability across
    different AQM algorithms but often underutilizes the link capacity.}
    \label{fig:throughput_no_burst_no_cross}
\end{figure}

\begin{figure}[t]
    \centering
    \includegraphics[width=0.9\linewidth]{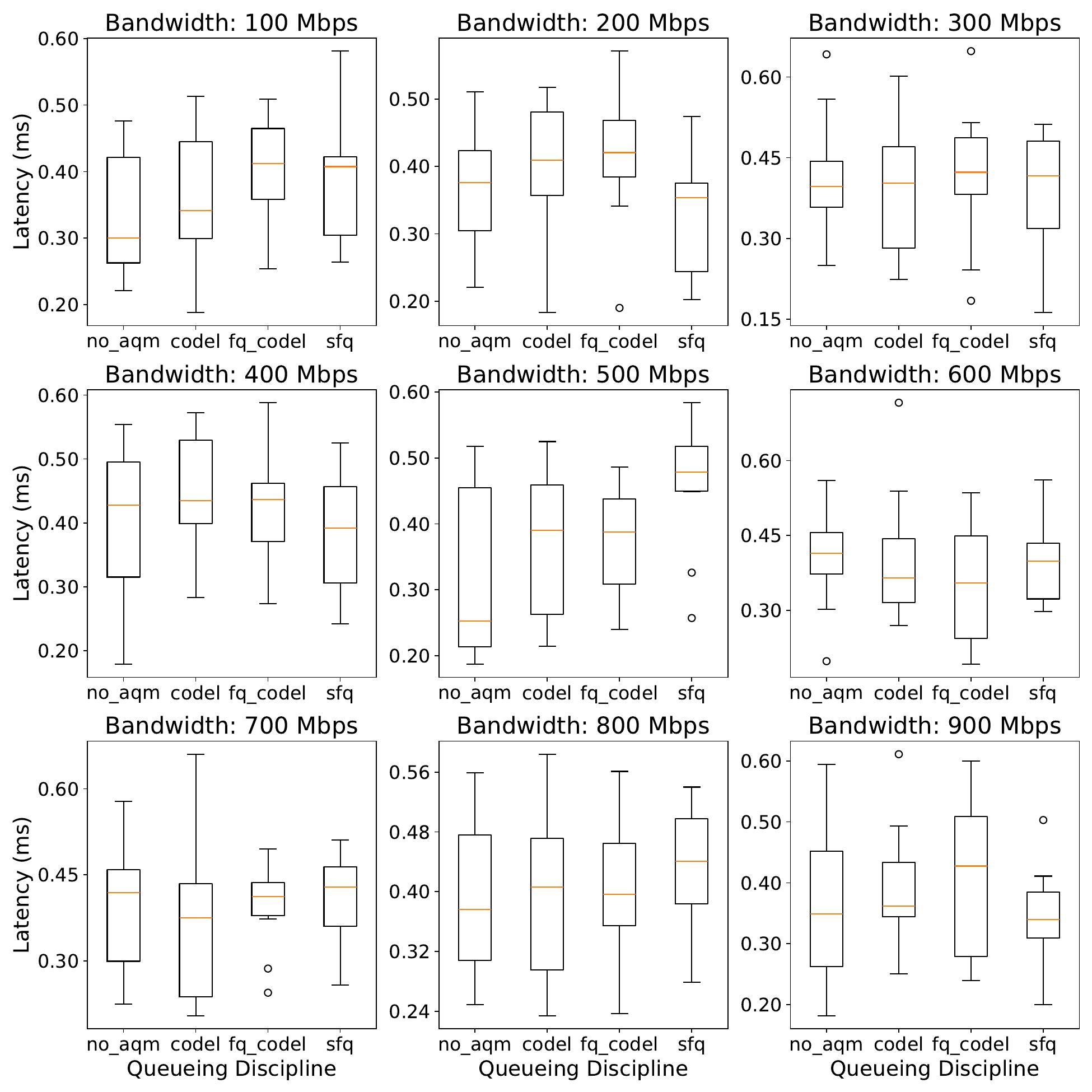}
    \tightcaption{Latency measurements without burst shaping and without cross traffic does not introduce significant latency spikes due 
    to low of queuing delays.}
    \label{fig:latency_no_burst_no_cross}
\end{figure}

\subsection{End-to-end measurement results reported with NDT}

For all the results presented in this section, we run NDT speed tests on our lab setup as introduced in the Section~\ref{sec:methodology}.
We measure speed tests across all AQM algorithms and every test is repeated 10 times. We refer to the AQM methods with the labels 
\texttt{no\_aqm}, \texttt{codel}, \texttt{fq\_codel}, and \texttt{sfq} for the No AQM, CoDel, FQ-CoDel, and SFQ AQM algorithms respectively.
In our terminology, \texttt{no\_aqm} refers to \texttt{pfifo} which is the default drop-tail queuing discipline in Linux.

\mypara{NDT speed tests on the lab testbed without burst shaping or TCP Cubic cross traffic}
We use our lab setup as introduced in the Section~\ref{sec:methodology} to run NDT speed tests
without explicitly adding cross traffic. This represents a baseline scenario where the only traffic is from the speed test
measurement tool and the measurements are not performed under load. We do not introduce any link latency using \texttt{netem}
for this measurement. Figures \ref{fig:throughput_no_burst_no_cross} and \ref{fig:latency_no_burst_no_cross} show the throughput and latency measurements respectively.

We observe the distribution of the measurements have measureable variations across the AQM algorithms used, even if the mean values don't report 
larger variations. Furthermore, if we do not use burst shaping, we observe that the bandwidth of the shaped link can be 
significantly underutilized, especially for higher rate limits (\eg At bandwidth of 800 Mbps, the maximum link utilization is less than 700 Mbps) which 
affects the throughput measurements.

\mypara{NDT speed tests on the lab testbed with burst shaping without TCP Cubic cross traffic} 
For these measurements we run NDT speed tests on our lab setup by introducing burst shaping on the link but without adding any cross traffic.
This enables the NDT tool reach closer to the theoretical link rate capacity using traffic sent by the test itself is maximized, without needing added competing traffic to saturate the link. 

\begin{figure}[t]
    \centering
    \includegraphics[width=0.95\linewidth]{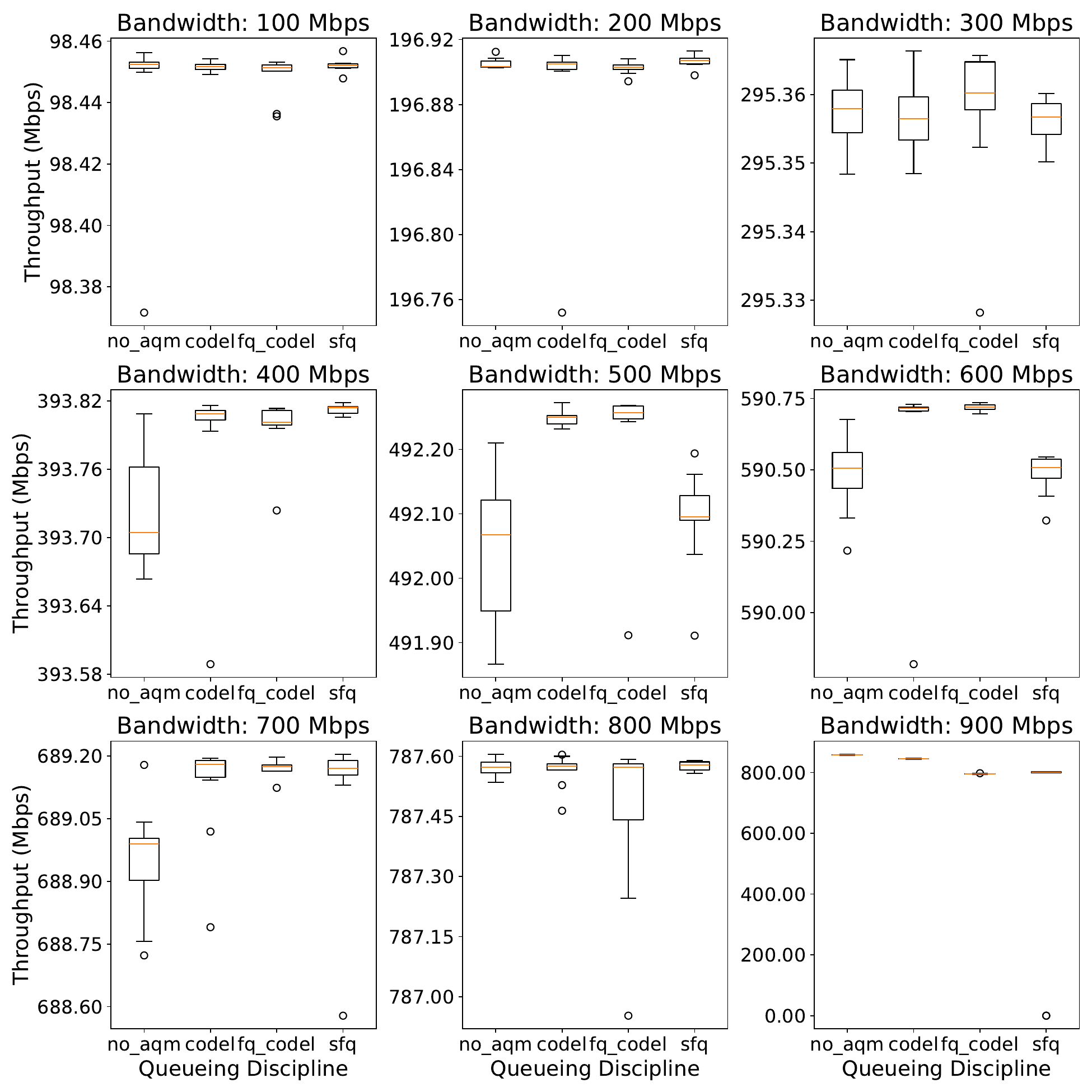}
    \tightcaption{Throughput measurement with burst shaping without cross traffic achieve higher link utilization and better use 
    of available bandwidth across all AQM algorithms.}
    \label{fig:throughput_with_burst_no_cross}
\end{figure}

\begin{figure}[t]
    \centering
    \includegraphics[width=0.95\linewidth]{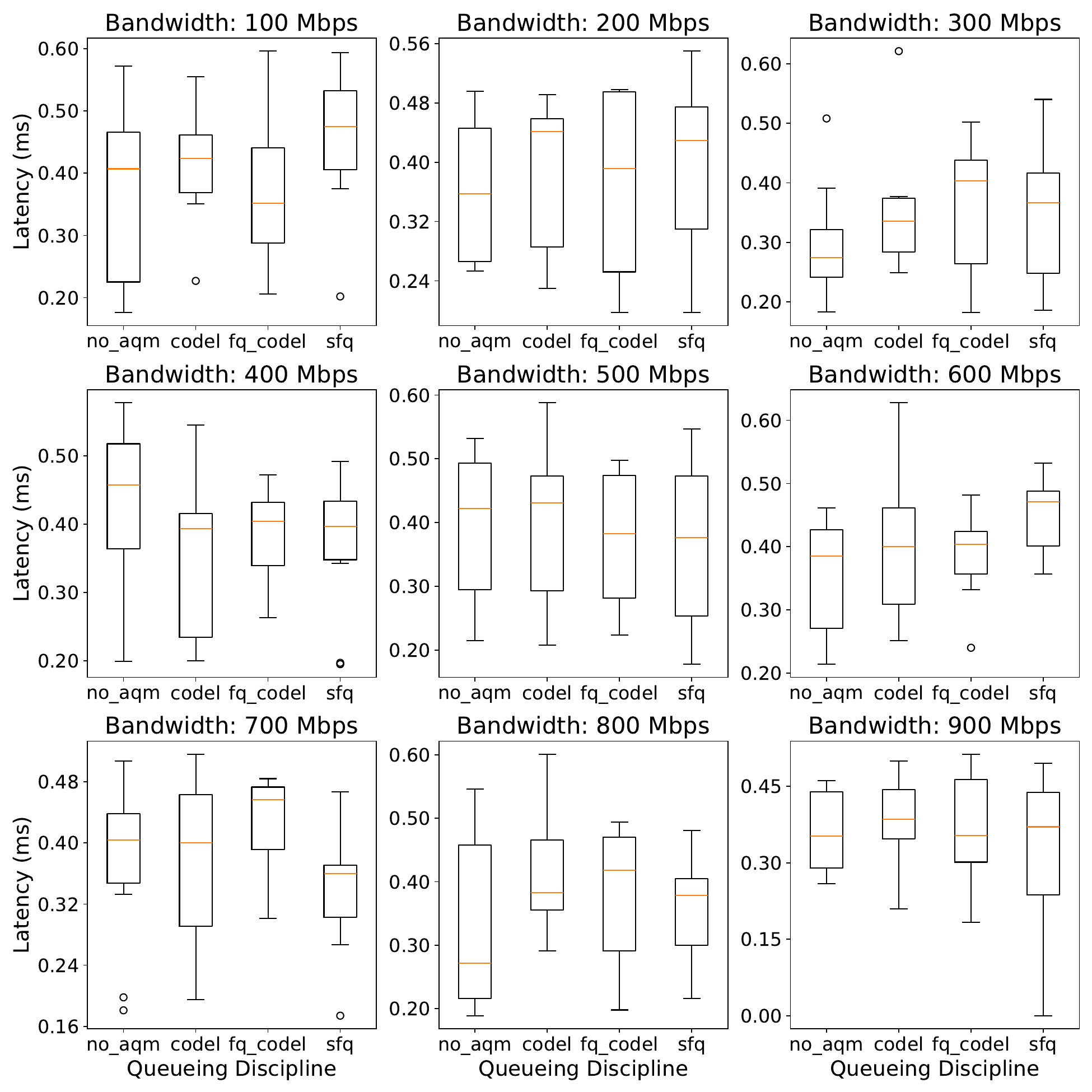}
    \tightcaption{Latency measurement with burst shaping without cross traffic is still stable due to lower competing load on the link.}
    \label{fig:latency_with_burst_no_cross}
\end{figure}

Figures \ref{fig:throughput_with_burst_no_cross} and \ref{fig:latency_with_burst_no_cross} show the throughput and latency measurements respectively.
We observe as opposed to the case without burst shaping, the throughput measurements are closer to the configured rate limits 
for the shaped link (\eg At bandwidth of 800 Mbps, the maximum link utilization is now close to 800 Mbps). We conclude burst allowance 
is necessary to allow speed test tools to send enough traffic to maximize link utilization and accurately measure throughput, especially at higher bandwidths.

\mypara{NDT speed tests on the lab testbed with burst shaping and TCP Cubic cross traffic} 
For these measurements we run NDT speed tests on our lab setup
with TCP Cubic as cross traffic. This represents a scenario where the speed test measurements are performed under load
from competing TCP traffic. TCP Cubic cross traffic is generated by running an \texttt{iperf3} test in parallel with the
NDT speed test as mentioned the section \ref{sec:methodology}.
The TCP flows are set to be capacity seeking and try to use all available bandwidth. We do not introduce any link latency using
\texttt{netem} on the server side. Finally, burst shaping is enabled on the link so that the speed test tool is able to
maximize the link utilization.

\begin{figure}[t]
    \centering
    \includegraphics[width=0.95\linewidth]{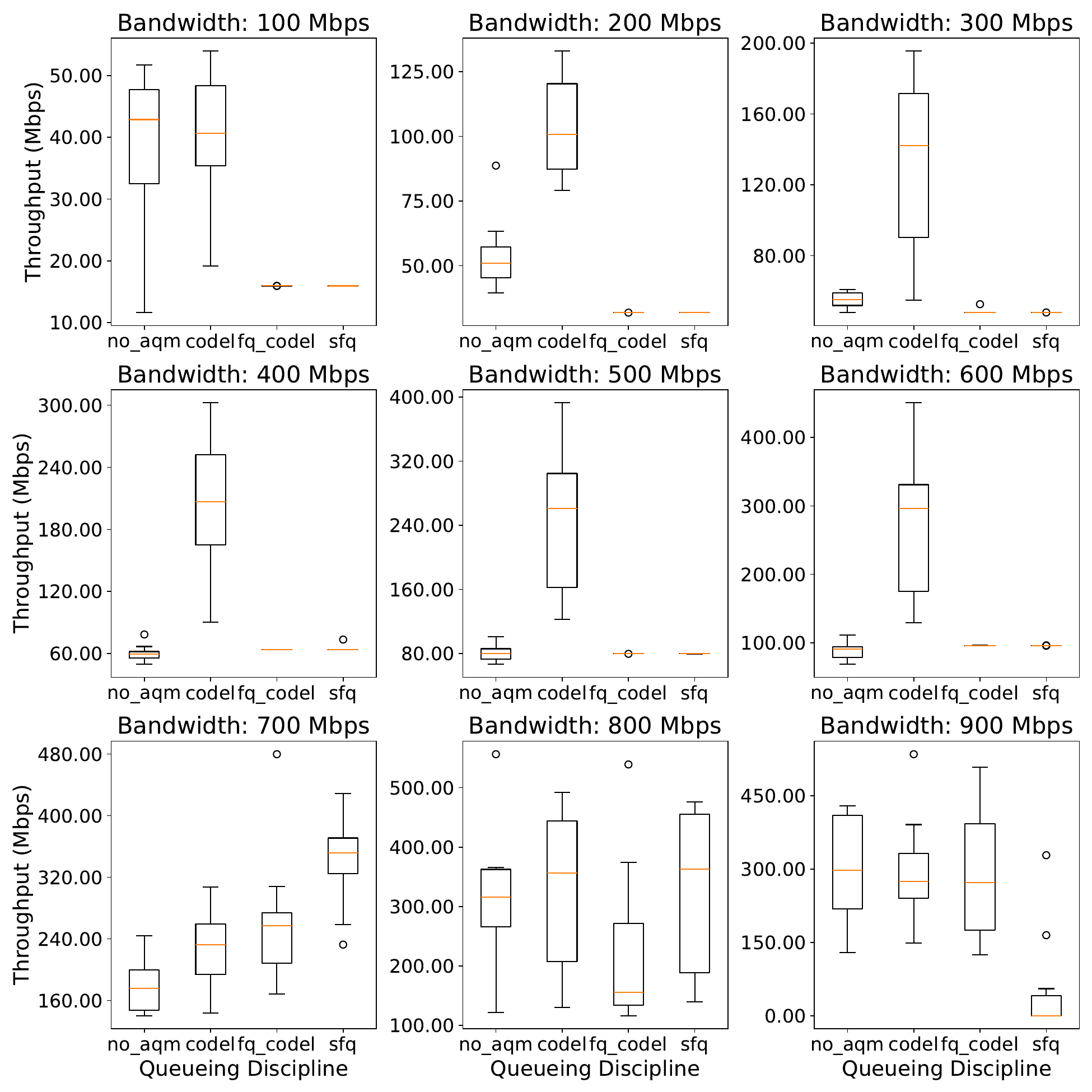}
    \tightcaption{Throughput measurement with burst shaping and TCP cross traffic shows higher variability across different AQM algorithms as 
    these become more prominent under load from competing traffic.}
    \label{fig:throughput_withcross}
\end{figure}

\begin{figure}[t]
    \centering
    \includegraphics[width=0.95\linewidth]{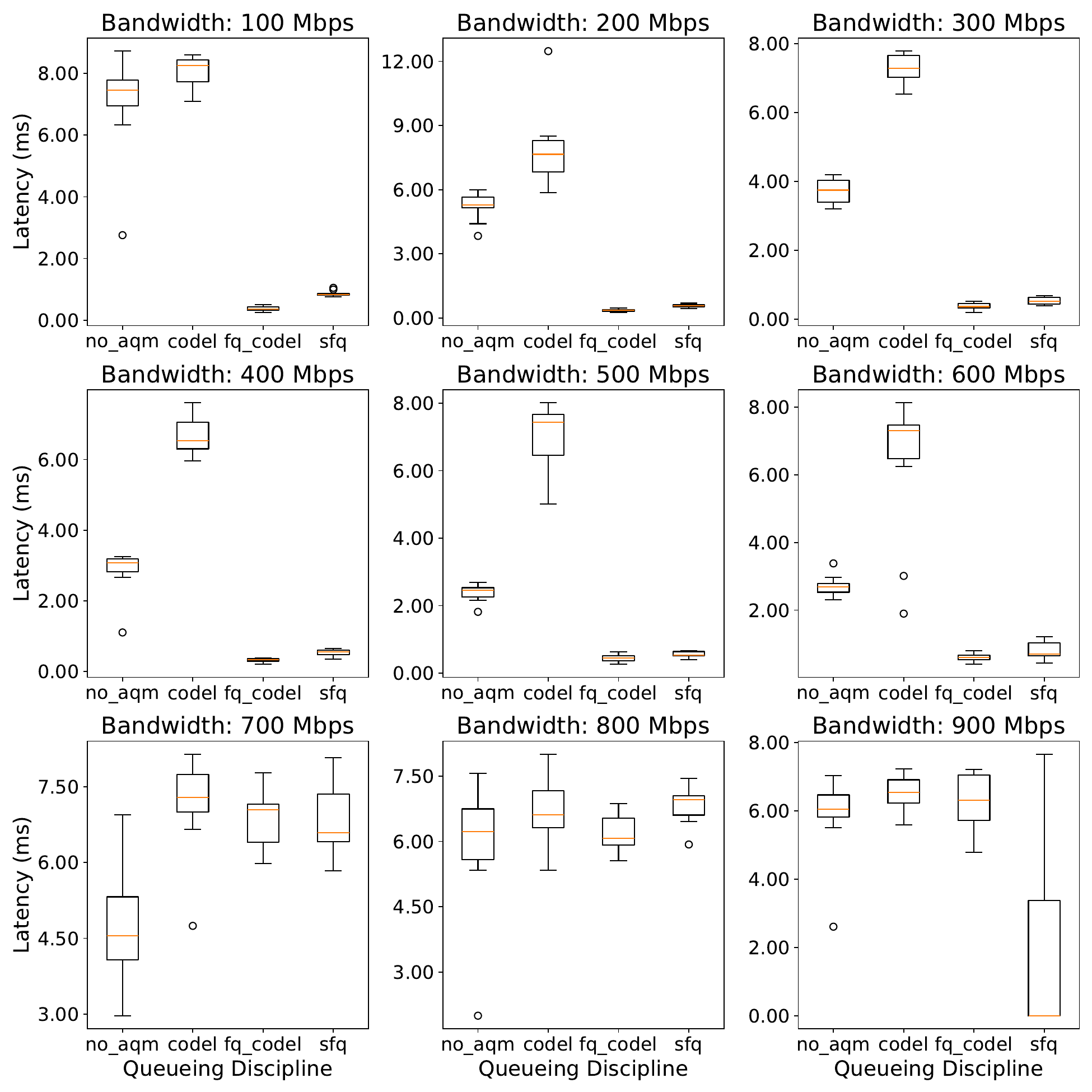}
    \caption{Latency measurement with burst shaping and TCP cross traffic shows significant increase in latency due to queuing delays
    introduced by competing traffic.}
    \label{fig:latency_withcross}
\end{figure}

Figure \ref{fig:throughput_withcross} shows that the throughput measurements are significantly lowered and variable
across all AQM algorithms as compared to scenarios without cross traffic (\eg (a) for \texttt{fq\_codel} at bandwidth 900 Mbps, the 
standard deviation is 138 Mbps with cross traffic and burst shaping but 1.51 Mbps without both and (b) cross traffic from the number of competing iperf3 TCP flows reduces the linkrate per flow for AQMs like \texttt{fq\_codel} due to its fariness policies, which causes significant variations even at small bandwidth increases of 100 Mbps). We also observe in Figure \ref{fig:latency_withcross} that the latency measurements are significantly higher
due to the queuing delays introduced by the cross traffic. We conclude that interpretation of speed test measurements 
by operators should consider the effect of network conditions and AQM policies in order to accurately interpret the results 
beyond a time-averaged value over the duration of the test.

\subsection{Instantaneous measurement results reported with NDT}

\begin{figure}[t]
    \centering
    \includegraphics[width=0.95\linewidth]{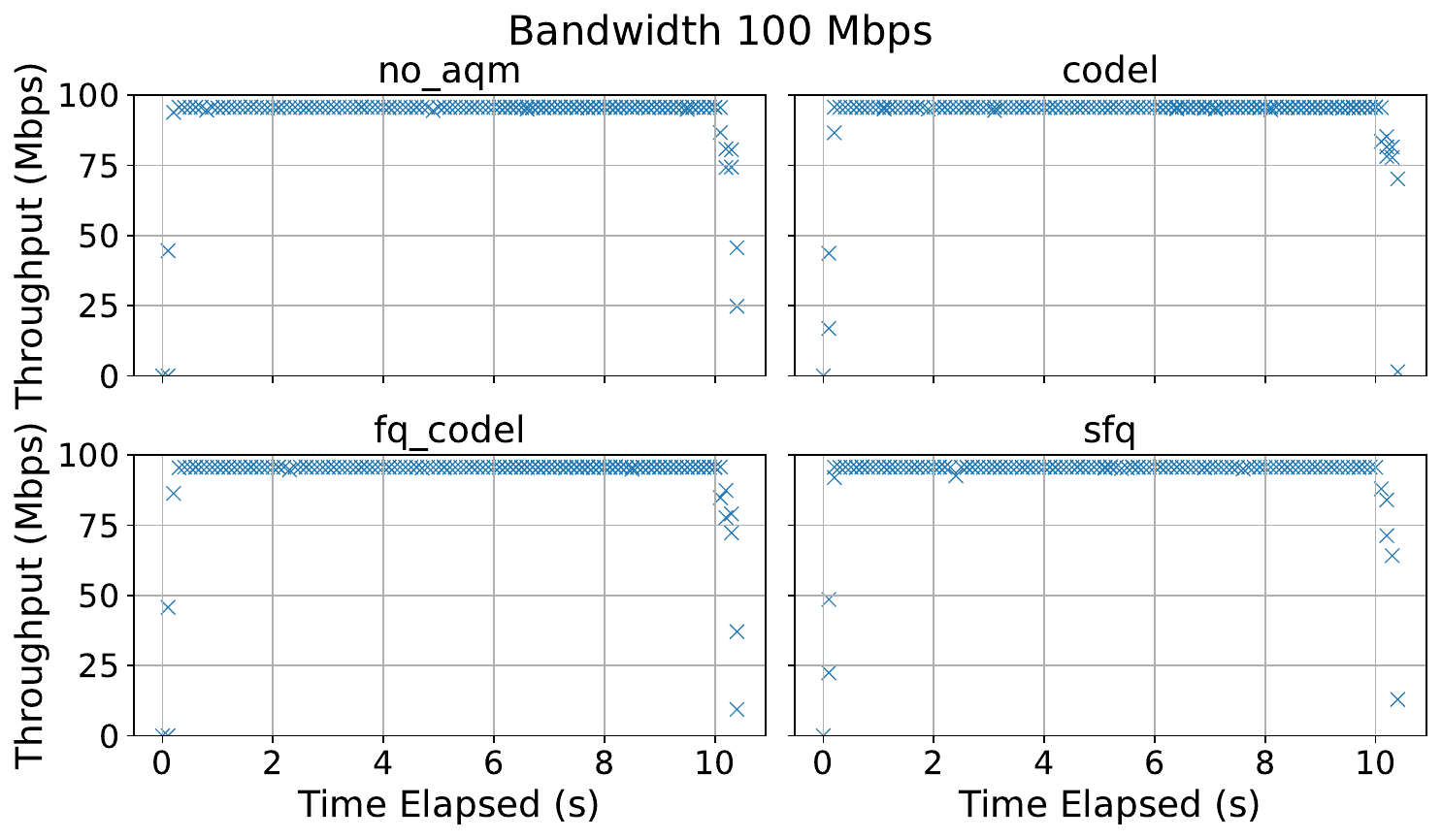}
    \tightcaption{Instantaneous upload throughput measurement without burst shaping or cross traffic at low bandwidth (100 Mbps) do not harness the 
     characteristics of different AQM policies.}
    \label{fig:instant_low_bw_base}
\end{figure}

\begin{figure}[t]
    \centering
    \includegraphics[width=0.95\linewidth]{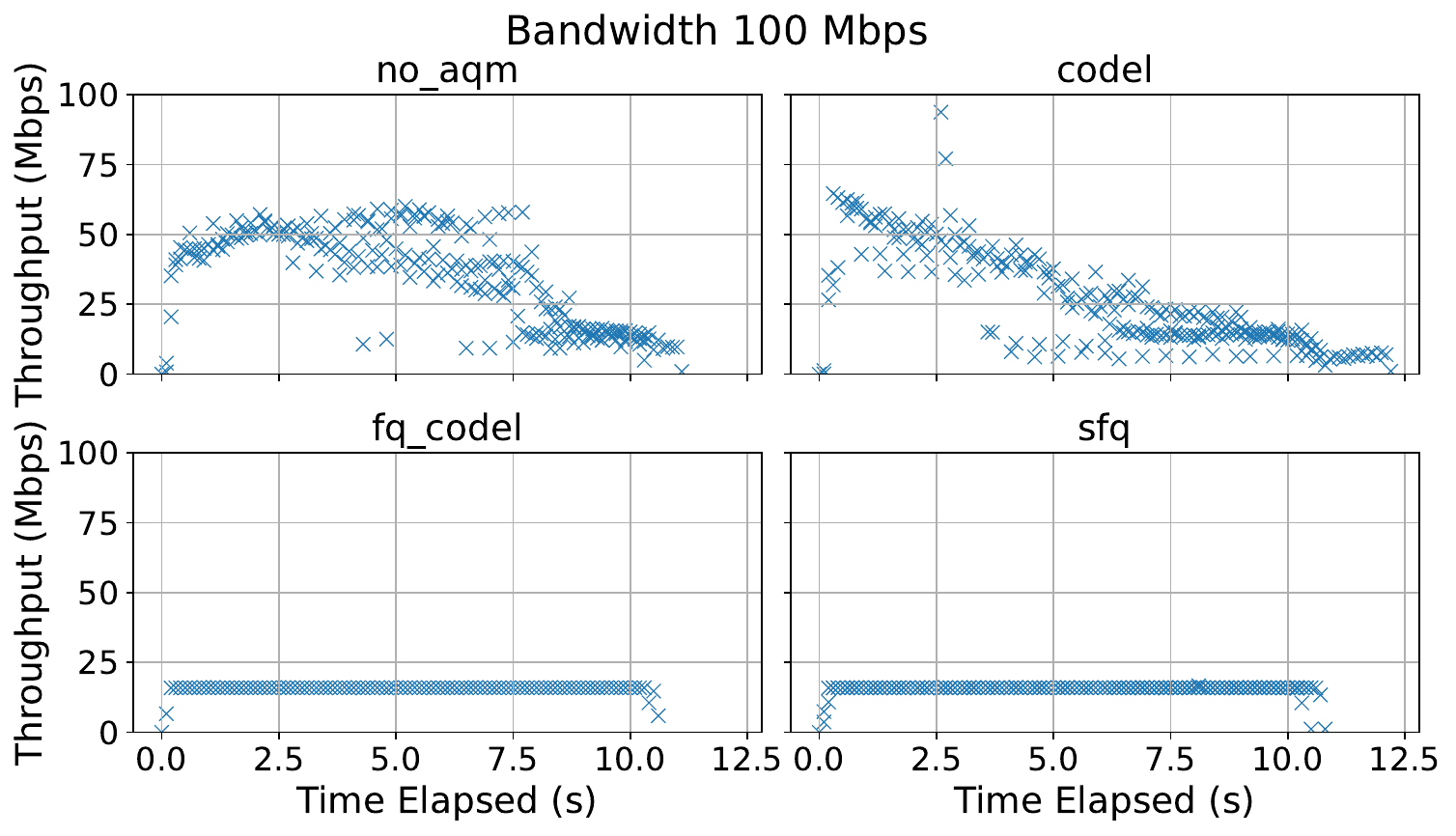}
    \caption{Instantaneous upload throughput measurement with burst shaping and cross traffic at low bandwidth (100 Mbps) shows some AQMs are 
    better at maintaining stable throughput.}
    \label{fig:instant_low_bw}
\end{figure}

\begin{figure}[t]
    \centering
    \includegraphics[width=0.95\linewidth]{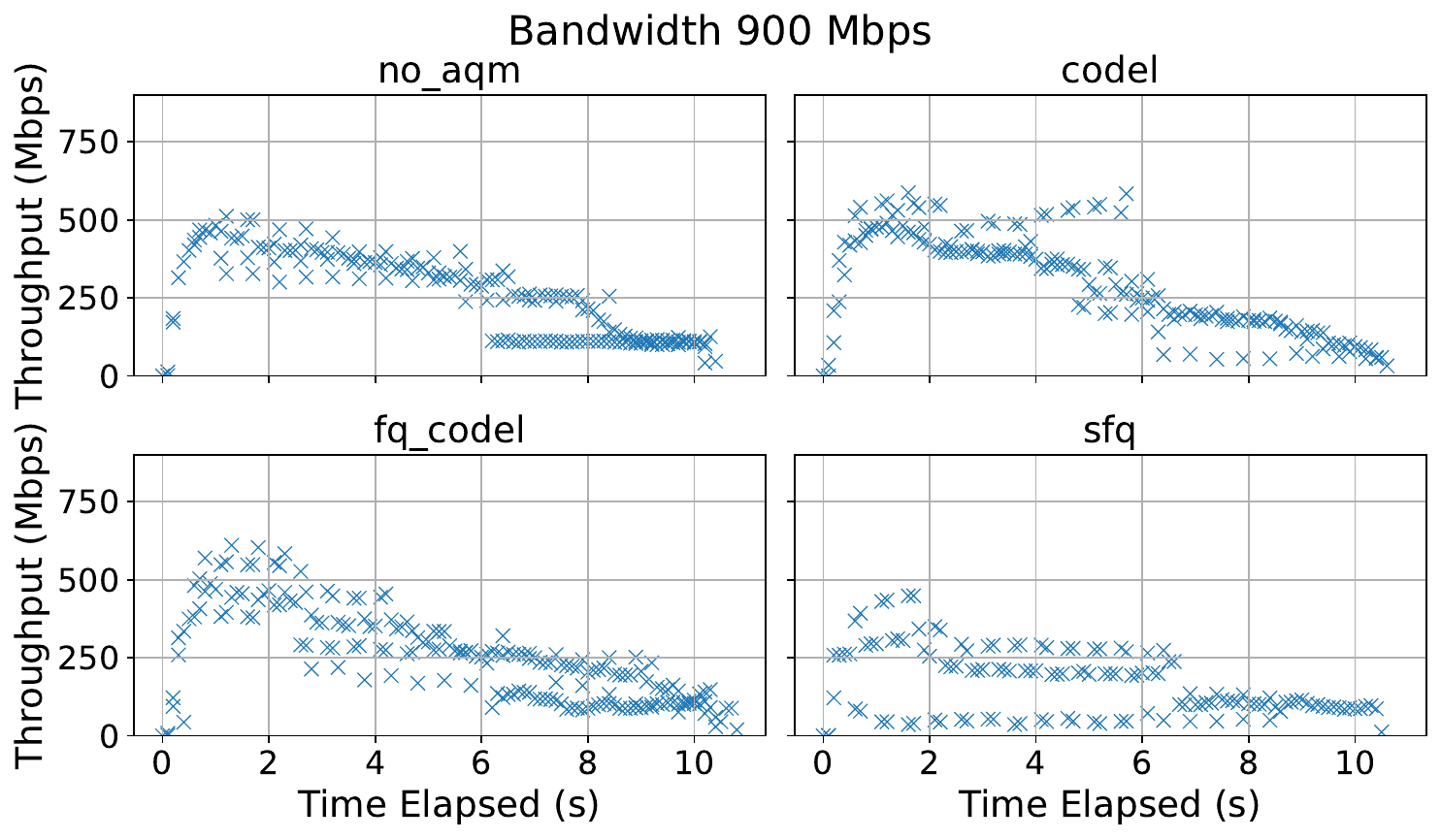}
    \tightcaption{Instantaneous upload throughput measurement with burst shaping and cross traffic at high bandwidth (900 Mbps) shows significant variability across 
    all AQM algorithms and affects the stability of throughput measurements.}
    \label{fig:instant_high_bw}
\end{figure}

We further study the variability of instantaneous throughput during the NDT speed tests by analyzing the time series data of upload throughput. 
This allows us to observe how throughput changes over time, particularly in response to network conditions and competing traffic.
We first consider a baseline scenario without burst shaping or cross traffic as shown in Figure \ref{fig:instant_low_bw_base}.
We see that the instantaneous throughput remains relatively stable across all of the AQM algorithms with minimal variability,
as also observed in Figure \ref{fig:throughput_no_burst_no_cross}. In measurements conducted under low load, 
the choice of AQM algorithm has minimal impact on throughput stability.

Next we analyze the instantaneous throughput with burst shaping and TCP Cubic cross traffic enabled. We compare two scenarios: one with low bandwidth limit (100 Mbps) and another with high bandwidth limit (900 Mbps). In both scenarios, 
we enable burst shaping and introduce TCP Cubic cross traffic.
We see at lower bandwidth limits, the instantaneous throughput remains relatively stable across two of AQM algorithms (\texttt{fq\_codel} and \texttt{sfq}) as shown in Figure \ref{fig:instant_low_bw}.
However, at higher bandwidth limits, the instantaneous throughput exhibits significant variability across all AQMs as shown in Figure \ref{fig:instant_high_bw}. 
Hence, measuring the impact of load and AQM policies on the stability and variability of speed test tools is
crucial for accurate interpretation of the results.

\section{Conclusion}
\label{sec:conclusion}

We present a measurement study analyzing the impact of AQM algorithms on speed test measurement tools, focusing on 
variability across different AQM policies and network conditions. We show that speed test measurements can exhibit significant variability 
across test duration and reporting only aggregated values misses important dynamics of network performance.
Hence, interpretation of speed test measurements should be aware of these underlying network factors while the 
tests themselves may be evolved to report more fine-grained metrics.

\bibliographystyle{ACM-Reference-Format}
\bibliography{refs}

@misc{midoglu2018monroenettestconfigurabletooldissecting,
  title         = {MONROE-Nettest: A Configurable Tool for Dissecting Speed Measurements in Mobile Broadband Networks},
  author        = {Cise Midoglu and Leonhard Wimmer and Andra Lutu and Ozgu Alay and Carsten Griwodz},
  year          = {2018},
  eprint        = {1710.07805},
  archiveprefix = {arXiv},
  primaryclass  = {cs.NI},
  url           = {https://arxiv.org/abs/1710.07805}
}

@inproceedings{bauer2010understanding,
  title        = {Understanding broadband speed measurements},
  author       = {Bauer, Steven and Clark, David D and Lehr, William},
  year         = {2010},
  organization = {TPRC}
}

@misc{feamster2019internetspeedmeasurementcurrent,
  title         = {Internet Speed Measurement: Current Challenges and Future Recommendations},
  author        = {Nick Feamster and Jason Livingood},
  year          = {2019},
  eprint        = {1905.02334},
  archiveprefix = {arXiv},
  primaryclass  = {cs.NI},
  url           = {https://arxiv.org/abs/1905.02334}
}

@misc{CeroWRT_speedtests,
  author       = {{Dave Täht}},
  title        = {The new features and flaws of speedtest.net, ookla and cloudflare},
  howpublished = {\url{https://blog.cerowrt.org/post/speedtests/}},
  note         = {Accessed: 2025-10-01}
}

@techreport{ietf-ippm-responsiveness-07,
  number      = {draft-ietf-ippm-responsiveness-07},
  type        = {Internet-Draft},
  institution = {Internet Engineering Task Force},
  publisher   = {Internet Engineering Task Force},
  note        = {Work in Progress},
  url         = {https://datatracker.ietf.org/doc/draft-ietf-ippm-responsiveness/07/},
  author      = {Christoph Paasch and Randall Meyer and Stuart Cheshire and Will Hawkins},
  title       = {{Responsiveness under Working Conditions}},
  pagetotal   = 41,
  year        = 2025,
  month       = jul,
  day         = 7
}

@article{hoilandbufferbloat,
  title={Bufferbloat and Beyond},
  author={H{\o}iland-J{\o}rgensen, Toke}
}

@article{bufferbloat,
author = {Gettys, Jim and Nichols, Kathleen},
title = {Bufferbloat: Dark Buffers in the Internet: Networks without effective AQM may again be vulnerable to congestion collapse.},
year = {2011},
issue_date = {November 2011},
publisher = {Association for Computing Machinery},
address = {New York, NY, USA},
volume = {9},
number = {11},
issn = {1542-7730},
url = {https://doi.org/10.1145/2063166.2071893},
doi = {10.1145/2063166.2071893},
journal = {Queue},
month = {nov},
pages = {40–54},
numpages = {15}
}

@misc{l4s,
      title={Dual Queue Coupled AQM: Deployable Very Low Queuing Delay for All}, 
      author={Koen De Schepper and Olga Albisser and Olivier Tilmans and Bob Briscoe},
      year={2022},
      eprint={2209.01078},
      archivePrefix={arXiv},
      primaryClass={cs.NI},
      url={https://arxiv.org/abs/2209.01078}, 
}

@article{red,
  author={Floyd, S. and Jacobson, V.},
  journal={IEEE/ACM Transactions on Networking}, 
  title={Random early detection gateways for congestion avoidance}, 
  year={1993},
  volume={1},
  number={4},
  pages={397-413},
  doi={10.1109/90.251892}}

@article{codel,
author = {Nichols, Kathleen and Jacobson, Van},
title = {Controlling queue delay},
year = {2012},
issue_date = {July 2012},
publisher = {Association for Computing Machinery},
address = {New York, NY, USA},
volume = {55},
number = {7},
issn = {0001-0782},
url = {https://doi.org/10.1145/2209249.2209264},
doi = {10.1145/2209249.2209264},
journal = {Commun. ACM},
month = {jul},
pages = {42–50},
numpages = {9}
}

@misc{fqcodel,
    series =    {Request for Comments},
    number =    8290,
    howpublished =  {RFC 8290},
    publisher = {RFC Editor},
    doi =       {10.17487/RFC8290},
    url =       {https://www.rfc-editor.org/info/rfc8290},
    author =    {Toke Høiland-Jørgensen and Paul McKenney and dave.taht@gmail.com and Jim Gettys and Eric Dumazet},
    title =     {{The Flow Queue CoDel Packet Scheduler and Active Queue Management Algorithm}},
    pagetotal = 25,
    year =      2018,
    month =     jan,
}

@conference{sfq
    ,author="Paul E. McKenney"
    ,title="Stochastic Fairness Queuing"
    ,Year="1990"
    ,Month="June"
    ,booktitle="IEEE INFOCOM'90 Proceedings"
    ,publisher="The Institute of Electrical and Electronics Engineers, Inc."
    ,pages="733-740"
    ,address="San Francisco"
}

@inproceedings{drr,
author = {Shreedhar, M. and Varghese, George},
title = {Efficient fair queueing using deficit round robin},
year = {1995},
isbn = {0897917111},
publisher = {Association for Computing Machinery},
address = {New York, NY, USA},
url = {https://doi.org/10.1145/217382.217453},
doi = {10.1145/217382.217453},
booktitle = {Proceedings of the Conference on Applications, Technologies, Architectures, and Protocols for Computer Communication},
pages = {231–242},
numpages = {12},
location = {Cambridge, Massachusetts, USA},
series = {SIGCOMM '95}
}

@INPROCEEDINGS{pie,
  author={Pan, Rong and Natarajan, Preethi and Piglione, Chiara and Prabhu, Mythili Suryanarayana and Subramanian, Vijay and Baker, Fred and VerSteeg, Bill},
  booktitle={2013 IEEE 14th International Conference on High Performance Switching and Routing (HPSR)}, 
  title={PIE: A lightweight control scheme to address the bufferbloat problem}, 
  year={2013},
  volume={},
  number={},
  pages={148-155},
  doi={10.1109/HPSR.2013.6602305}}

@misc{cake,
      title={Piece of CAKE: A Comprehensive Queue Management Solution for Home Gateways}, 
      author={Toke Høiland-Jørgensen and Dave Täht and Jonathan Morton},
      year={2018},
      eprint={1804.07617},
      archivePrefix={arXiv},
      primaryClass={cs.NI},
      url={https://arxiv.org/abs/1804.07617}, 
}

@misc{ndt,
  title        = {Network Diagnostic Tool (NDT)},
  author       = {{Measurement Lab}},
  howpublished = {\url{https://www.measurementlab.net/tests/ndt/}},
  year         = {2025},
  note         = {Accessed: 2025-10-02}
}

@misc{ndt7spec,
  title        = {NDT7 Protocol Specification},
  author       = {{Measurement Lab}},
  howpublished = {\url{https://github.com/m-lab/ndt-server/blob/main/spec/ndt7-protocol.md}},
  year         = {2025},
  note         = {Accessed: 2025-10-02}
}

@misc{ookla,
  title        = {Speedtest by Ookla},
  author       = {{Ookla}},
  howpublished = {\url{https://www.speedtest.net/}},
  year         = {2025},
  note         = {Accessed: 2025-10-02}
}

@misc{ookla_method,
  title        = {Understanding Speedtest Methodology},
  institution  = {Ookla},
  howpublished = {\url{https://www.ookla.com/articles/ookla-speedtest-methodology}},
  year         = {2023},
  note         = {Whitepaper, Accessed: 2025-10-02}
}

@misc{apple_speedtest,
  title        = {Apple Speedtest},
  author       = {{Apple Inc.}},
  howpublished = {\url{https://support.apple.com/HT212313}}, 
  year         = {2025},
  note         = {Accessed: 2025-10-02}
}

@misc{fast,
  title        = {Fast.com Internet Speed Test},
  author       = {{Netflix}},
  howpublished = {\url{https://fast.com/}},
  year         = {2025},
  note         = {Accessed: 2025-10-02}
}

@inproceedings{clark2021measurement,
title     = {Measurement, meaning and purpose: Exploring the M-Lab NDT dataset},
author    = {Clark, David D and Wedeman, Sara},
booktitle = {TPRC49: The 49th Research Conference on Communication, Information and Internet Policy},
year      = {2021},
url       = {https://dx.doi.org/10.2139/ssrn.3898339}
}

@inproceedings{clark2024measurement,
title     = {Measurement of Internet access latency: A cross-dataset comparison},
author    = {Clark, David D and Wedeman, Sara},
booktitle = {Proceedings of the TPRC2024 The Research Conference on Communications, Information and Internet Policy},
year      = {2024},
url       = {http://dx.doi.org/10.2139/ssrn.4909679}
}

@misc{crpa2019broadband,
title        = {Broadband Availability and Access},
author       = {{Center for Rural Pennsylvania}},
year         = {2018},
url          = {https://www.rural.pa.gov/publications/broadband.cfm},
note         = {Accessed November 13, 2024},
organization = {Center for Rural Pennsylvania},
type         = {Web Page}
}

@article{cdcc2022illustrating,
title   = {Illustrating Internet Speed Divides in the Caribbean During COVID-19},
author  = {Alexander, Dale and Døhl Diouf, Lika},
journal = {FOCUS Magazine of the Caribbean Development and Cooperation Committee (CDCC)},
year    = {2022},
month   = {12},
type    = {Report},
url     = {https://hdl.handle.net/11362/48956}
}

@inproceedings{hoiland-jorgensen2016measuring,
author    = {H\o{}iland-J\o{}rgensen, Toke and Ahlgren, Bengt and Hurtig, Per and Brunstrom, Anna},
title     = {Measuring Latency Variation in the Internet},
year      = {2016},
isbn      = {9781450342926},
publisher = {Association for Computing Machinery},
address   = {New York, NY, USA},
url       = {https://doi.org/10.1145/2999572.2999603},
doi       = {10.1145/2999572.2999603},
booktitle = {Proceedings of the 12th International on Conference on Emerging Networking EXperiments and Technologies},
pages     = {473-480},
numpages  = {8},
location  = {Irvine, California, USA},
series    = {CoNEXT '16}
}

@misc{jiang2023mobile,
title         = {Mobile Internet Quality Estimation using Self-Tuning Kernel Regression},
author        = {Hanyang Jiang and Henry Shaowu Yuchi and Elizabeth Belding and Ellen Zegura and Yao Xie},
year          = {2023},
eprint        = {2311.05641},
archiveprefix = {arXiv},
primaryclass  = {stat.AP},
url           = {https://arxiv.org/abs/2311.05641}
}

@misc{lee2023analyzing,
title         = {Analyzing Disparity and Temporal Progression of Internet Quality through Crowdsourced Measurements with Bias-Correction},
author        = {Hyeongseong Lee and Udit Paul and Arpit Gupta and Elizabeth Belding and Mengyang Gu},
year          = {2023},
eprint        = {2310.16136},
archiveprefix = {arXiv},
primaryclass  = {stat.AP},
url           = {https://arxiv.org/abs/2310.16136}
}

@article{macmillan2023comparative,
author     = {MacMillan, Kyle and Mangla, Tarun and Saxon, James and Marwell, Nicole P. and Feamster, Nick},
title      = {A Comparative Analysis of Ookla Speedtest and Measurement Labs Network Diagnostic Test (NDT7)},
year       = {2023},
issue_date = {March 2023},
publisher  = {Association for Computing Machinery},
address    = {New York, NY, USA},
volume     = {7},
number     = {1},
url        = {https://doi.org/10.1145/3579448},
doi        = {10.1145/3579448},
journal    = {Proc. ACM Meas. Anal. Comput. Syst.},
month      = mar,
articleno  = {19},
numpages   = {26}
}

@inproceedings{nabi2024red,
author    = {Nabi, Syed Tauhidun and Wen, Zhuowei and Ritter, Brooke and Hasan, Shaddi},
title     = {Red is Sus: Automated Identification of Low-Quality Service Availability Claims in the US National Broadband Map},
year      = {2024},
isbn      = {9798400705922},
publisher = {Association for Computing Machinery},
address   = {New York, NY, USA},
url       = {https://doi.org/10.1145/3646547.3688441},
doi       = {10.1145/3646547.3688441},
booktitle = {Proceedings of the 2024 ACM on Internet Measurement Conference},
pages     = {2-18},
numpages  = {17},
location  = {Madrid, Spain},
series    = {IMC '24}
}

@inproceedings{paul2022characterizing,
author    = {Paul, Udit and Liu, Jiamo and Farias-llerenas, David and Adarsh, Vivek and Gupta, Arpit and Belding, Elizabeth},
title     = {Characterizing Internet Access and Quality Inequities in California M-Lab Measurements},
year      = {2022},
isbn      = {9781450393478},
publisher = {Association for Computing Machinery},
address   = {New York, NY, USA},
url       = {https://doi.org/10.1145/3530190.3534813},
doi       = {10.1145/3530190.3534813},
booktitle = {Proceedings of the 5th ACM SIGCAS/SIGCHI Conference on Computing and Sustainable Societies},
pages     = {257-265},
numpages  = {9},
location  = {Seattle, WA, USA},
series    = {COMPASS '22}
}

@inproceedings{paul2022importance,
author    = {Paul, Udit and Liu, Jiamo and Gu, Mengyang and Gupta, Arpit and Belding, Elizabeth},
title     = {The importance of contextualization of crowdsourced active speed test measurements},
year      = {2022},
isbn      = {9781450392594},
publisher = {Association for Computing Machinery},
address   = {New York, NY, USA},
url       = {https://doi.org/10.1145/3517745.3561441},
doi       = {10.1145/3517745.3561441},
booktitle = {Proceedings of the 22nd ACM Internet Measurement Conference},
pages     = {274-289},
numpages  = {16},
location  = {Nice, France},
series    = {IMC '22}
}

@article{sanchez-arias2023understanding,
author   = {Sanchez-Arias, Reinaldo and Jaimes, Luis G. and Taj, Shahram and Habib, Md. Selim},
journal  = {IEEE Access},
title    = {Understanding the State of Broadband Connectivity: An Analysis of Speedtests and Emerging Technologies},
year     = {2023},
volume   = {11},
number   = {},
pages    = {101580-101603},
doi      = {10.1109/ACCESS.2023.3313231},
url      = {https://ieeexplore.ieee.org/abstract/document/10244030}
}

@article{saxon2022what,
title    = {What we can learn from selected, unmatched data: Measuring internet inequality in Chicago},
journal  = {Computers, Environment and Urban Systems},
volume   = {98},
pages    = {101874},
year     = {2022},
issn     = {0198-9715},
doi      = {https://doi.org/10.1016/j.compenvurbsys.2022.101874},
url      = {https://www.sciencedirect.com/science/article/pii/S0198971522001181},
author   = {James Saxon and Dan A. Black}
}

@article{sharma2024beyond,
author     = {Sharma, Taveesh and Schmitt, Paul and Bronzino, Francesco and Feamster, Nick and Marwell, Nicole P.},
title      = {Beyond Data Points: Regionalizing Crowdsourced Latency Measurements},
year       = {2024},
issue_date = {December 2024},
publisher  = {Association for Computing Machinery},
address    = {New York, NY, USA},
volume     = {8},
number     = {3},
url        = {https://doi.org/10.1145/3700416},
doi        = {10.1145/3700416},
journal    = {Proc. ACM Meas. Anal. Comput. Syst.},
month      = dec,
articleno  = {34},
numpages   = {24}
}

@inproceedings{sundaresan2017challenges,
author    = {Sundaresan, Srikanth and Deng, Xiaohong and Feng, Yun and Lee, Danny and Dhamdhere, Amogh},
title     = {Challenges in inferring internet congestion using throughput measurements},
year      = {2017},
isbn      = {9781450351188},
publisher = {Association for Computing Machinery},
address   = {New York, NY, USA},
url       = {https://doi.org/10.1145/3131365.3131382},
doi       = {10.1145/3131365.3131382},
booktitle = {Proceedings of the 2017 Internet Measurement Conference},
pages     = {43-56},
numpages  = {14},
location  = {London, United Kingdom},
series    = {IMC '17}
}

@inproceedings{ndt_mlab_pam2010,
  author    = {Sundaresan, Srikanth and de Donato, Walter and Feamster, Nick and Teixeira, Renata and Crawford, Sam and Pescapè, Antonio},
  title     = {Broadband Internet performance: A view from the edge},
  booktitle = {Proceedings of the Passive and Active Measurement Conference (PAM)},
  year      = {2011},
  pages     = {9--22},
  publisher = {Springer},
  doi       = {10.1007/978-3-642-19260-9_2}
}

@inproceedings{ndt_mlab_sigcomm2018,
  author    = {Bozakov, Zdravko and Schulman, Aaron and Sundaresan, Srikanth},
  title     = {M-Lab: An Open Platform for Large-Scale Network Measurement},
  booktitle = {Proceedings of the ACM SIGCOMM Conference (Demo Session)},
  year      = {2018},
  pages     = {80--81},
  publisher = {ACM},
  doi       = {10.1145/3234200.3234239}
}

@inproceedings{broadband_pam2019,
  author    = {Bischof, Zachary S. and Callado, Andr{\'e} and Sundaresan, Srikanth and Feamster, Nick},
  title     = {Characterizing latency in broadband access networks},
  booktitle = {Proceedings of the Passive and Active Measurement Conference (PAM)},
  year      = {2019},
  pages     = {65--77},
  publisher = {Springer},
  doi       = {10.1007/978-3-030-15986-3_5}
}

@inproceedings{bottleneck_imc2020,
  author    = {Rao, Aditya and Gupta, Arpit and Feamster, Nick and Krishnamurthy, Balachander},
  title     = {Understanding Bottlenecks in the Internet Last Mile},
  booktitle = {Proceedings of the ACM Internet Measurement Conference (IMC)},
  year      = {2020},
  pages     = {321--335},
  publisher = {ACM},
  doi       = {10.1145/3419394.3423642}
}

@inproceedings{ookla_sigmetrics2016,
  author    = {Arzani, Babak and Benson, Theophilus and Maltz, David and Popa, Lucian},
  title     = {Speedtest at Scale: Measurement and Biases in Crowdsourced Performance Data},
  booktitle = {Proceedings of the ACM SIGMETRICS Conference},
  year      = {2016},
  pages     = {307--318},
  publisher = {ACM},
  doi       = {10.1145/2896377.2901480}
}

@inproceedings{fast_imc2022,
  author    = {Smith, Jacob and Sundaresan, Srikanth and Scott, Will and Eravuchira, Siddharth and Feamster, Nick},
  title     = {Understanding Netflix's Fast.com: A Study of Web-Based Speed Measurement},
  booktitle = {Proceedings of the ACM Internet Measurement Conference (IMC)},
  year      = {2022},
  pages     = {458--470},
  publisher = {ACM},
  doi       = {10.1145/3517745.3561442}
}

@inproceedings{crowd_pam2021,
  author    = {Restrepo, Juan and Rula, John and Dainotti, Alberto},
  title     = {Crowdsourced Internet Measurement: Bias and Representativeness},
  booktitle = {Proceedings of the Passive and Active Measurement Conference (PAM)},
  year      = {2021},
  pages     = {123--136},
  publisher = {Springer},
  doi       = {10.1007/978-3-030-72582-2_8}
}

@article{floyd2001tcp,
author = {Floyd, S.},
title = {A report on recent developments in TCP congestion control},
year = {2001},
issue_date = {April 2001},
publisher = {IEEE Press},
volume = {39},
number = {4},
issn = {0163-6804},
url = {https://doi.org/10.1109/35.917508},
doi = {10.1109/35.917508},
journal = {Comm. Mag.},
month = apr,
pages = {84--90},
numpages = {7}
}

@misc{tc_manpage,
  title = {tc - traffic control},
  author = {{Linux man-pages project}},
  organization = {{The Linux Kernel}},
  note = {Accessed through `man tc` on a Linux system},
  year = {2025},
  url = {https://man7.org/linux/man-pages/man8/tc.8.html}
}

@inproceedings{10.1007/978-3-031-85960-1_10,
author="Sarpkaya, Fatih Berkay
and Fund, Fraida
and Panwar, Shivendra",
editor="Testart, Cecilia
and van Rijswijk-Deij, Roland
and Stiller, Burkhard",
title="To Adopt or Not to Adopt L4S-Compatible Congestion Control? Understanding Performance in a Partial L4S Deployment",
booktitle="Passive and Active Measurement",
year="2025",
publisher="Springer Nature Switzerland",
address="Cham",
pages="217--246",
isbn="978-3-031-85960-1"
}

@techreport{ietf-tsvwg-l4sops-08,
    number =    {draft-ietf-tsvwg-l4sops-08},
    type =      {Internet-Draft},
    institution =   {Internet Engineering Task Force},
    publisher = {Internet Engineering Task Force},
    note =      {Work in Progress},
    url =       {https://datatracker.ietf.org/doc/draft-ietf-tsvwg-l4sops/08/},
    author =    {Greg White},
    title =     {{Operational Guidance on Coexistence with Classic ECN during L4S Deployment}},
    pagetotal = 23,
    year =      2025,
    month =     jul,
    day =       7,
}

@misc{iperf3,
  author = {Dugan, Jon and Elliott, Seth and Mah, Bruce A. and Poskanzer, Jeff and Prabhu, Kaustubh},
  title = {{iPerf3: A TCP, UDP, and SCTP network bandwidth measurement tool}},
  year = {2014},
  publisher = {ESnet},
  url = {https://iperf.fr/},
  note = {Accessed: [Insert Date Here]}
}

\appendix
\section{Ethics Statement}

This research was conducted in a lab environment using our own testbed infrastructure.
All experiments were performed on a private network with no involvement of human subjects or measurements from 
external users. This work does not pose any ethical or privacy concerns.

\section{Code Snippets for Measurement Setup}

\subsection{Code Snippets for data collection with NDT7 and remote tcpdump}

\noindent
\begin{minipage}{\columnwidth}
\begin{lstlisting}[style=mypython,
  caption={Initializing file paths and creating the remote output directory.},
  label={lst:ndt_setup}
]
def run_speedtest_with_remote_pcap(output_root, server_ip):
    """
    Runs ndt7-client locally while capturing PCAP remotely via SSH.
    """
    current_time = time.strftime("%Y%m%d-%H%M%S")
    json_file = f"{output_root}/speedtest_{current_time}.jsonl"
    remote_pcap = f"/home/noise/{output_root}/speedtest_{current_time}.pcap"

    print(f"[*] Creating remote directory: {output_root}")
    mkdir_cmd = f"mkdir -p {output_root}"
    mkdir_result = ssh_command_server(mkdir_cmd)
    if mkdir_result.returncode != 0:
        print(f"[!] Failed to create remote directory {output_root}")
        return None, None

    print(f"[*] Remote PCAP will be saved to: {remote_pcap}")
\end{lstlisting}
\end{minipage}

\noindent
\begin{minipage}{\columnwidth}
\begin{lstlisting}[style=mypython,
  caption={Starting remote tcpdump using adapter\_unsynced and background execution.},
  label={lst:ndt_start_tcpdump}
]
    tcpdump_remote_cmd = (
        f"tcpdump -j adapter_unsynced -w {remote_pcap} host {server_ip} "
        "> /dev/null 2>&1 & echo $!"
    )

    print("[*] Starting remote tcpdump...")
    start_proc = ssh_command_server(tcpdump_remote_cmd)

    if start_proc.returncode != 0:
        print("[!] Failed to start remote tcpdump:")
        print(start_proc.stderr)
        return None, None

    tcpdump_pid = start_proc.stdout.strip()
    print(f"[*] Remote tcpdump started with PID {tcpdump_pid}")
\end{lstlisting}
\end{minipage}

\noindent
\begin{minipage}{\columnwidth}
\begin{lstlisting}[style=mypython,
  caption={Running the NDT7 client locally and stopping the remote capture.},
  label={lst:ndt_run_stop}
]
    try:
        os.system(
            f"ndt7-client -server {server_ip} -no-verify "
            f"-format json > {json_file}"
        )
    finally:
        print("[*] Stopping remote tcpdump...")
        ssh_command_server(f"sudo kill -SIGINT {tcpdump_pid}")

    print(f"Speedtest JSON saved locally at: {json_file}")
    print(f"PCAP saved remotely at: {remote_pcap}")

    return json_file, remote_pcap
\end{lstlisting}
\end{minipage}

\subsection{Helper script for SSH Commands}

\noindent
\begin{minipage}{\columnwidth}
\begin{lstlisting}[style=mypython,
  caption={Helper function to run a command via SSH on the remote server.},
  label={lst:ssh_command}
]
def ssh_command_server(cmd):
    key_path = "/path/.ssh/id_rsa"  # adjust to the one from ssh -v
    server = "<name>@" + server_ip.split(":")[0]
    result = subprocess.run(
        ["ssh", "-o", "StrictHostKeyChecking=no", "-i", key_path, server, cmd],
        stdout=subprocess.PIPE,
        stderr=subprocess.PIPE,
        text=True,
    )
    return result

def scp_command_server(remote_path, local_path):
    key_path = "/path/.ssh/id_rsa"  # adjust to the one from ssh -v
    server = "<name>@" + server_ip.split(":")[0]
    result = subprocess.run(
        ["scp", "-o", "StrictHostKeyChecking=no", "-i", key_path, f"{server}:{remote_path}", local_path],
        stdout=subprocess.PIPE,
        stderr=subprocess.PIPE,
        text=True,
    )
    return result
\end{lstlisting}
\end{minipage}

\noindent
\begin{minipage}{\columnwidth}
\begin{lstlisting}[style=mypython,
  caption={Helper function to run a command via SSH on the remote server.},
  label={lst:scp_command}
]
def collect_remote_pcap(remote_pcap_path, local_output_dir):
    """
    Copies the remote PCAP file back to local storage.
    """
    remote_pcap_path = remote_pcap_path.strip()
    local_path = os.path.join(local_output_dir, os.path.basename(remote_pcap_path))

    print(f"[*] Fetching {remote_pcap_path} from remote host...")
    print(f"[*] Saving to local path: {local_path}")
    result = scp_command_server(remote_pcap_path, local_path)
    if result.returncode != 0:
        print("[!] Failed to fetch remote PCAP:")
        print(result.stderr)
        return None

    print(f" PCAP saved locally to {local_path}")
    return local_path
\end{lstlisting}
\end{minipage}

\subsection{Code Snippet for AQM shaping with tc}

\noindent
\begin{minipage}{\columnwidth}
\begin{lstlisting}[style=mybash,
  caption={Bash script header, variable setup, and \texttt{start()} function.},
  label={lst:tc_start}
]
#!/bin/bash
CMD=$1
IFACE=$2
BW=$3
LATENCY=$4
LOSS=$5
AQM_METHOD=$6

TC=/sbin/tc

start() {
    if [[ -z "$BW" || -z "$LATENCY" || -z "$LOSS" ]]; then
        echo "Usage: $0 start <interface> <bandwidth_kbps> <latency_ms> <loss_percentage> <aqm_method>"
        exit 1
    fi

    echo "Applying traffic shaping with AQM ($AQM_METHOD) to $IFACE..."

    $TC qdisc del dev $IFACE root 2>/dev/null
    $TC qdisc add dev $IFACE root handle 1: htb default 11
    $TC class add dev $IFACE parent 1: classid 1:11 htb rate ${BW}kbit \
        ceil $((BW * 16 / 10))kbit burst $((BW * 36 / 1000))k \
        cburst $((BW * 36 / 1000))k
\end{lstlisting}
\end{minipage}

\noindent
\begin{minipage}{\columnwidth}
\begin{lstlisting}[style=mybash,
  caption={Applying the selected AQM or tail drop in \texttt{start()}.},
  label={lst:tc_aqm}
]
    if [[ "$AQM_METHOD" == "no_aqm" ]]; then
        $TC qdisc add dev $IFACE parent 1:11 handle 20: pfifo limit 100
        echo "Traffic shaping applied with tail drop (pfifo) on $IFACE"
    else
        $TC qdisc add dev $IFACE parent 1:11 handle 20: ${AQM_METHOD}
        echo "Traffic shaping applied with $AQM_METHOD on $IFACE"
    fi
}
\end{lstlisting}
\end{minipage}

\noindent
\begin{minipage}{\columnwidth}
\begin{lstlisting}[style=mybash,
  caption={Stop function, show function, and case statement handling commands.},
  label={lst:tc_stop_show}
]
stop() {
    echo "Removing traffic shaping from $IFACE..."
    $TC qdisc del dev $IFACE root 2>/dev/null
}

show() {
    echo "Current traffic control settings for $IFACE:"
    $TC qdisc show dev $IFACE
    $TC class show dev $IFACE
    $TC filter show dev $IFACE

    AQM_CHECK=$($TC -s qdisc show dev $IFACE | grep -E "fq_codel|cake|pie|sfq|codel")
    if [[ -n "$AQM_CHECK" ]]; then
        echo " AQM ($AQM_METHOD) is active on $IFACE."
    elif [[ "$AQM_METHOD" == "no_aqm" ]]; then
        echo " Tail drop (pfifo) is applied instead of AQM on $IFACE."
    else
        echo " AQM is NOT applied!"
    fi
}

case "$CMD" in
    start)
        start
        show
        ;;
    stop)
        stop
        ;;
    show)
        show
        ;;
    *)
        echo "Usage: $0 {start|stop|show} <interface> <bandwidth_kbps> <latency_ms> <loss_percentage> <aqm_method>"
        ;;
esac
\end{lstlisting}
\end{minipage}

\end{document}